\documentclass[aps,prl,twocolumn,floatfix,superscriptaddress,longbibliography]{revtex4}
\usepackage{bm}
\usepackage{epsf}
\usepackage{amssymb}
\usepackage{amsmath}
\usepackage{graphicx}
\usepackage{rotating}
\usepackage{epsfig}
\usepackage{psfrag}
\usepackage{amsmath}
\usepackage{hyperref}
\usepackage{subfigure}
\usepackage{bm}% bold math
\usepackage{color}

\renewcommand{\a}{\alpha}
\renewcommand{\b}{\beta}

\renewcommand{\d}{\delta}
\renewcommand{\S}{\Sigma}

\newcommand{\D}{\Delta}
\renewcommand{\th}{\theta}

\renewcommand{\o}{\omega}
\renewcommand{\O}{\Omega}
\newcommand{\e}{\epsilon}
\renewcommand{\l}{\lambda}

\renewcommand{\dag}{\dagger}

\newcommand{\nn}{\nonumber}

\renewcommand{\dag}{\dagger}

\def\mr{\mathrm}
\def\mc{\mathcal}

\newcommand{\ket}[1]{\left|{#1}\right\rangle}

\newcommand{\be}{\begin{equation}}
\newcommand{\ee}{\end{equation}}
\newcommand{\ba}{\begin{eqnarray}}
\newcommand{\ea}{\end{eqnarray}}

\newcommand {\apgt} {\ {\raise-.5ex\hbox{$\buildrel>\over\sim$}}\ }
\newcommand {\aplt} {\ {\raise-.5ex\hbox{$\buildrel<\over\sim$}}\ }

\begin{document}

\title{Variable-range hopping through marginally localized phonons}

\author{Sumilan Banerjee}
\affiliation{Department of Condensed Matter Physics, Weizmann Institute of Science, Rehovot, 7610001, Israel}
\author{Ehud Altman}
\affiliation{Department of Condensed Matter Physics, Weizmann Institute of Science, Rehovot, 7610001, Israel}

\begin{abstract}
We investigate the effect of coupling Anderson localized particles in one dimension to a system of marginally localized phonons having a symmetry protected delocalized mode at zero frequency. This situation is naturally realized for electrons coupled to phonons in a disordered nano-wire as well as for ultra-cold fermions coupled to phonons of a superfluid  in a one dimensional disordered trap. To determine if the coupled system can be many-body localized we analyze  the phonon-mediated hopping transport for both the weak and strong coupling regimes. We show that the usual variable-range hopping mechanism involving a low-order phonon processes is ineffective at low temperature due to discreteness of the bath at the required energy. Instead, the system thermalizes through a many-body process  involving exchange of a diverging number  $n\propto -\log T$ of phonons in the low temperature limit. This effect leads to a highly singular prefactor to Mott's well known formula and strongly suppresses the variable range hopping rate. Finally we comment on possible implications of this physics in higher dimensional electron-phonon coupled systems.
\end{abstract}

\maketitle

%----------------------------------------------------------------------------------------------------------------------------------
%\section{Introduction}

Recently there has been considerable interest in the phenomenon of many-body localization (MBL) in disordered systems \cite{Gornyi2005,Basko2006} (and see \cite{Nandkishore2015,Altman2015} for reviews of recent work), whereby quantum matter can avoid thermal equilibrium and exhibit persistent quantum correlations in dynamics \cite{Bahri2015,Serbyn2014}. Experimental investigations of MBL have also been recently reported in systems of ultra-cold atoms \cite{Schreiber2015} as well as trapped ion chains \cite{Smith2015}.

The need for decoupling from the environment, however, inhibits realization of true MBL in conventional solids where electrons  are coupled to a bath of delocalized phonon modes, though possible precursors of such localization have been seen \cite{Ovadia2015}. 
Electron-phonon interaction leads to a transport mechanism known as hopping conductivity \cite{Miller1960,Shklovskii1984}. At low-temperature and neglecting electron-electron interactions,  phonon assisted hopping leads to well-known Mott's law \cite{Mott1968} of variable range hopping (VRH) rate, $\tau^{-1}\sim \exp [-(T_0/T)^{1/(d+1)}]$ . 

The VRH mechanism relies on the ability of the electrons to draw energy from the phonons to hop from one localized state to another and  that the phonons constitute a `good' bath, having a continuous density of states (DOS). In low dimensions, however, this bath can be fragile as the phonons themselves get localized. Here we discuss the effect of phonon localization on VRH transport and whether there is a possibility of many-body localization (MBL) of the coupled electron-phonon (el-ph) system.

The natural physical realizations of phonons in low dimensions have very different localization properties than electrons. One such realization involves phonons of a disordered nano-wire, which can be modeled as a harmonic chain of random masses coupled by random springs. Another realization involves phonons of a superfluid in a disordered potential. In both cases the phonons are Goldstone modes associated with spontaneous breaking of a continuous symmetry: translational symmetry in the random solid and $U(1)$ charge symmetry in the superfluid. The symmetry protects a delocalized mode at zero frequency corresponding to a uniform transformation of the entire system. In one and two dimensional systems phonons at all non-vanishing frequencies are localized with a localization length that diverges on approaching the zero frequency limit \cite{John1983,Gurarie2003}. The question we ask is whether such phonons, in spite of being only asymptotically delocalized, can facilitate delocalization of another species of particles (e.g. electrons) coupled to them.

To answer this question we analyze the phonon mediated hopping rate of electrons for both weak and strong electron-phonon coupling. We find that the conventional VRH transport involving one- or a few-phonon process is not effective at low temperatures in a one dimensional system. Nonetheless, the electrons do ultimately delocalize via increasingly high-order processes in electron-phonon coupling, which involve $n\propto-\ln(T/T_0)$ phonons  as $T\to 0$. This leads to a modified VRH rate with a highly singular pre-exponential factor. We also consider the situation in two dimensions and show that in this case the usual VRH rate is operative, at least for weak disorder. %We discuss the implications of our results for generic inelastic processes in the context of MBL in interacting quantum systems. Below we give a outline of the paper.

We note that Potter and Nandkishore discussed a somewhat similar question, concerning MBL in a system of interacting particles having asymptotically delocalized single particle states \cite{Nandkishore2014}. That paper is relevant for example to Chern insulators, where a delocalized single particle state necessarily separates between localized bands with different Chern indices. In such systems a particle can delocalize already through a fourth order process involving virtual excitation to a nearly delocalized state of the particle. The difference from our result, which requires arbitrarily high order in the low temperature limit, stems from the fact that in the electron-phonon system the electron delocalizes  by coupling to a secondary particle. There are no delocalized electronic states to virtually excite. 

In Ref. \cite{Nandkishore2015a} Nandkishore considered the coupling of a localized system to a weak bath showing that the bath could either lead to delocalization of the particles or alternatively become localized itself. However, for the phonon bath we consider, as already mentioned, the zero frequency mode is symmetry protected against localization and can not be `proximity localized'.   Huse et. al. \cite{Huse2015}  on the other hand considered coupling to a delocalized bath albeit of finite size (i.e. one featuring discrete spectrum with Wigner-Dyson level statistics). Such a bath is again different from the phonon bath and is ineffective in delocalizing strongly localized particles. Finally we note that the phonon system acts as a dynamical quantum bath for the electrons. This is very different from an external classical drive as considered for example in the context of periodically driven Floquet systems \cite{Roy2015,Lazarides2015}.

Before analyzing a concrete model in detail we first discuss the conventional variable range hopping scheme and show how it fails for phonons in the random chain. In a phonon assisted hopping process an electron is transferred from a  single-particle localized state $\ket{l}$ centered at site $R_l$ to the localized state $\ket{l'}$ together with emission or absorption of a phonon that bridges the energy difference $\epsilon_l-\epsilon_{l'}$ between the two electronic states. The transition matrix element is proportional to the spatial overlap of the electronic states and hence decays exponentially with the range of the hop $R$. 

The hopping rate out of a localized electronic state can be obtained through Fermi's golden rule as:
\begin{equation}
\frac{1}{\tau}\simeq g^2\sum_R  e^{-\Delta_R/T} \,e^{-R/\xi}\, \nu(\Delta_R).\label{eq.FGRRate}
\end{equation}
Here $\Delta_R\simeq \Delta_\xi (\xi/R)^d$ is the typical energy offset to the nearest electronic state localized within a range $R$ of the initial state. $\Delta_\xi$ is the energy level spacing within a localization volume $~\xi^d$; $\Delta_R$ is the energy a thermal phonon has to provide. Hence the Boltzmann factor $e^{-\Delta_R/T}$, valid at low temperatures $T\ll \Delta_\xi$ ($k_\mr{B}=1$). Finally $\nu(\omega)$ is the phonon density of states and $g$ the strength of the el-ph coupling. Following Mott \cite{Mott1968}, the hopping range at a given temperature $R_M\approx \xi (\Delta_\xi/T)^{1/(d+1)}$, can be obtained from the saddle point of the sum in Eq.\eqref{eq.FGRRate}. Consequently, the electron draws an `optimal' energy $\Delta_M\approx \Delta_\xi (T/\Delta_\xi)^{d/(d+1)}$ from the phonon bath to make an incoherent hop. Replacing the sum with the saddle point values gives Mott's well known variable range hopping law
\ba
\frac{1}{\tau}\sim g^2\nu(\Delta_M) e^{-A_d\left(\Delta_\xi/T\right)^\frac{1}{d+1}}, \label{eq.VRH}
\ea
where $A_d$ in the exponent depends on $d$, e.g.~$A_1=2$. 

The fact that the phonons making up the bath may be localized was not taken into account in the simple Fermi golden rule calculation. As discussed in the introduction, for $d\leq 2$ the phonons are localized at all non-zero frequency $\omega$ with a localization length that diverges as $\omega\to 0$. Specifically in 1D the phonon localization length, $\ell_\omega\simeq \ell_0 (\omega_0/\omega)^\alpha$ with $1\leq\alpha\leq 2$ \cite{Gurarie2008}. Here the upper limit corresponds to weak disorder \cite{John1983,Gurarie2003}. The frequency $\omega_0$ is approximately the Debye frequency, related via the sound velocity $c$ to the microscopic length scale $\ell_0\simeq c/\omega_0$. Henceforth we set $\ell_0$ and $\o_0$ as the units of length and energy respectively (i.e. $\o_0=\ell_0=1$).
\begin{center}
\begin{figure}
\includegraphics[width=0.5\textwidth]{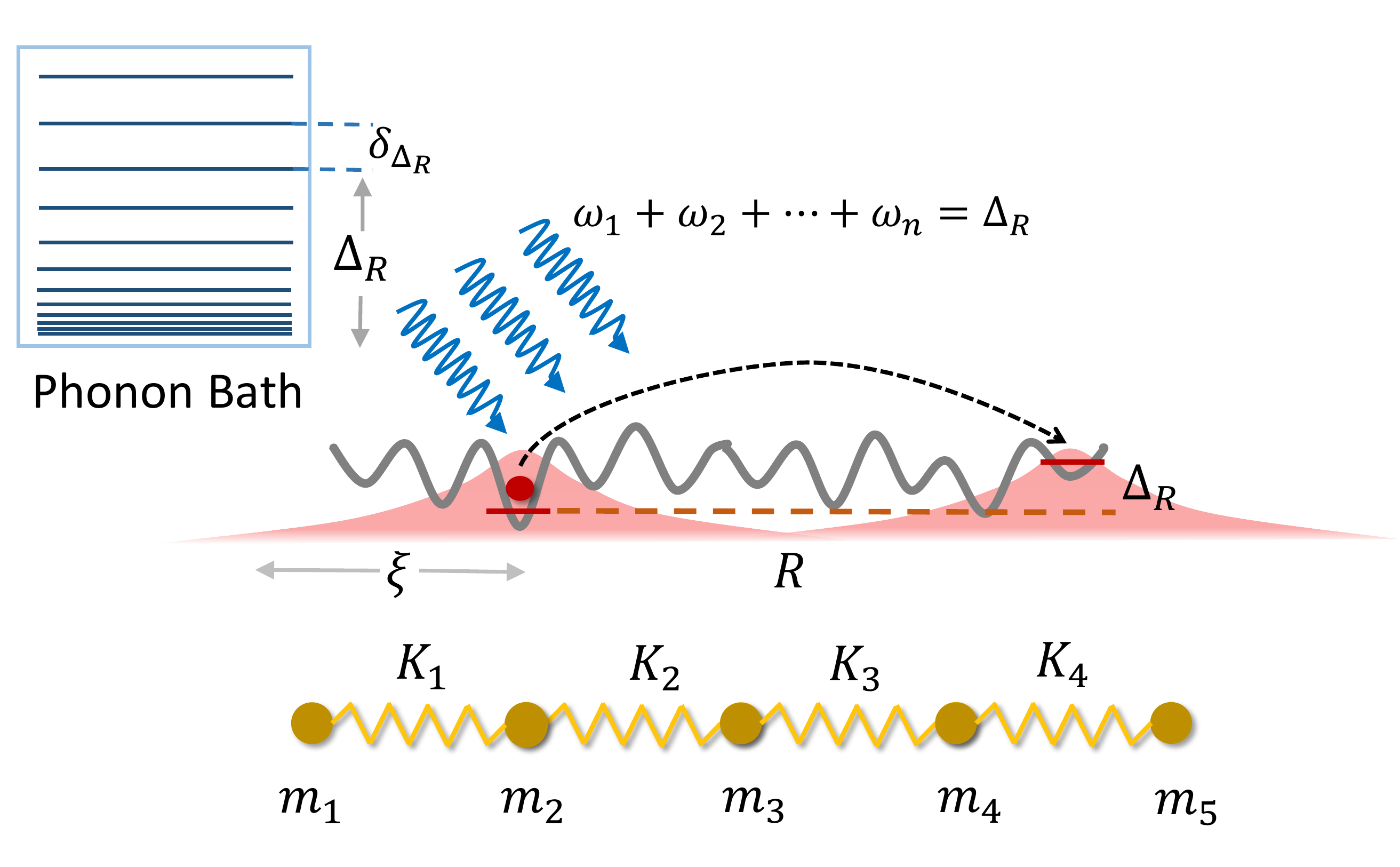}
\caption{The model of localized electron coupled to a disordered harmonic chain with random masses and spring constants. Electron with localization length $\xi$ may hop from one localized state to other due to exponential overlap $\sim \exp(-R/\xi)$ between the states by extracting an amount of energy $\D_R$ from the phonon-bath, e.g.~by absorbing a single phonon. However, due to the localized nature of the phonons, the one-phonon bath spectrum is discrete with a level spacing $\d_{\D_R}\simeq\D_R^\alpha$. The level spacing of the phonon bath rapidly collapses as $\D_R^{\alpha^n}$ for multi-phonon process involving $n$ phonons.}
\label{fig.Model}
\end{figure}
\end{center}

In the presence of phonon localization we can not assume a continuous density of states as done above. 
%The golden rule rate for hop between localized states in Eq.\eqref{eq.FGRRate} arises due to the lowest-order phonon process, i.e.~absorption or emission of a single phonon at energy $\Delta_R$. Since localized electrons have a discrete spectra, the effective `bath' density of states (DOS) entering in Eq.\eqref{eq.FGRRate} is that of a single phonon. 
Due to finite localization length, the `one-phonon bath' has a discrete DOS which only becomes continuous at $\omega\to 0$, as shown in Fig.~\ref{fig.Model}. The optimal hop that determines the VRH rate in Eq.~\eqref{eq.VRH} involves a phonon at energy $\Delta_M$, where the DOS is discrete with a level spacing $\delta_{\D_M}\simeq (\ell_{\Delta_M}\nu (\Delta_M))^{-1}\approx 1/\ell_{\Delta_M}$. Here we used the fact that the 1D phonon DOS remains constant $\nu(\o)\approx 1$ (in our units) at low frequencies for both weak and strong disorder \cite{Gurarie2003,Gurarie2008}. The golden rule rate is non-zero only if $\tau^{-1}>\delta_{\D_M}$ i.e. $g^2 \exp{(-2/\tilde{T}^{1/2})}> \Delta_\xi^\alpha\tilde{T}^{\alpha/2}$, where $\tilde{T}=T/\Delta_\xi$. The inequality is always violated as $T\to 0$. As a result, the VRH rate in Eq. \eqref{eq.VRH} is strictly zero at low temperature. 
%One can consider fluctuations around the optimal path in Eq.\eqref{eq.FGRRate} at the Gaussian level and obtain a modified rate $\propto  \exp{[-(2/\tilde{T}^{1/2})]}\tilde{T}^{1/4}$. This, however, does not alter the above conclusion about the vanishing of VRH rate of Eq.\eqref{eq.VRH}.

The question is naturally raised, whether the ineffectiveness of the usual VRH mechanism implies many-body localization of the electron-phonon system. Before coming to this conclusion, however, it is necessary to check if higher order phonon processes can lead to incoherent hopping. We now explain in simple terms the result, which is derived  in detail below.

We need to compare the putative decay rate of a localized state aided by multi-phonon emission or absorption with the relevant level spacing of such multi-phonon states. The transition matrix element is suppressed exponentially with the number ($n$) of emitted or absorbed phonons. We will show that this leads to a suppression of the standard VRH hopping rate by a factor $\sim \tilde{T}^{\frac{n\a -1}{2}}$. On the other hand by adding the discrete energies of $n$ phonons one can provide the required energy of the hop $\Delta_M$, up to a mismatch decreasing rapidly with $n$ as $\delta^{(n)}_{\D_M}\propto \Delta_M^{\alpha^n} \sim \tilde{T}^{\frac{\alpha^n}{2}}$   (see Supplementary Material).  Hence for a given temperature, the effective level spacing of a process involving more than $n_c\sim -\ln \tilde{T}$ phonons falls below the transition rate, thus allowing an incoherent transition, albeit with a highly suppressed rate. 

Having explained the essential arguments we now outline the explicit calculations using both a weak coupling and a strong coupling approach.  
%\section{Model} \label{sec:model}
We consider the following model
\ba
 \mc{H}&=&\mc{H}_{el}+\mc{H}_{ph}+\mc{H}_{el-ph}, \label{Eq.Model}
\ea
 $\mc{H}_{el}$ describes (spinless) electrons hopping with amplitude $t$ on a 1D lattice with on-site random potential $v_i$ at site $R_i$, i.e. $\mc{H}_{el}=-t\sum_{i} (c^\dagger_i c_{i+1}+ \mr{h.c.})+\sum_i v_i n_i$, with $n_i=c_i^\dagger c_i$. We write $\mc{H}_{el}=\sum_l \epsilon_l c_l^\dagger c_l$ in the basis of localized electronic states $\ket{l}$, i.e.~$\psi_l(R)\sim (e^{-|R-R_l|/2\xi}/\xi^{1/2})e^{i\theta_l(R)}$ with energy $\epsilon_l$. We retain a complex phase factor for the convenience of calculation. 

The part $\mc{H}_{ph}$ in Eq.\eqref{Eq.Model} describes the acoustic phonons of the random harmonic chain, i.e.
\ba
\mc{H}_{ph}&=&\sum_i \left[ \frac{p_i^2}{2m_i}+\frac{K_i}{2}(x_i-x_{i+1})^2\right]=\sum_\mu \omega_\mu a_\mu^\dagger a_\mu .~~ \label{eq.Hph}
\ea
where the masses $m_i$ and spring constants $K_i$'s are random such that the normal mode eigenfunction $\phi_\mu(R)\sim e^{-|R-R_\mu|/2\ell_\mu}/\ell_\mu^{1/2}$ are localized around $R=R_\mu$ with phonon localization length $\ell_\mu\sim \omega_\mu^{-\alpha}$. The phonon operators $a_\mu^\dagger$ and $a_\mu$ are related to the displacements, $x_i=\sum_\mu \phi_\mu(R_i)(a_\mu+a_\mu^\dagger)/\sqrt{2m_i\omega_\mu}$. The el-ph coupling term in Eq.\eqref{Eq.Model} is $\mc{H}_{el-ph}=\sum_{i,\mu} g_{\mu,i}n_i(a_\mu^\dag+a_\mu)$ where 
\ba
g_{\mu,i}\approx g\,\left(\frac{\o_\mu}{\ell_\mu}\right)^{1/2}e^{-|R_i-R_\mu|/2\ell_\mu}. \label{eq.el-phCoupling}
\ea
 Here $g$ parametrizes the coupling strength.

In the weak coupling approach we evaluate the hopping rate assisted by high-order (i.e. multiple) phonon processes using perturbation theory in the el-ph coupling $g$. Then, we compare the rate to the level spacing at each order in the expansion. 

%\subsection{Electronic self-energy and hopping rate} \label{Sec.SelfEnergy}
\begin{center}
\begin{figure}
\includegraphics[width=0.4\textwidth]{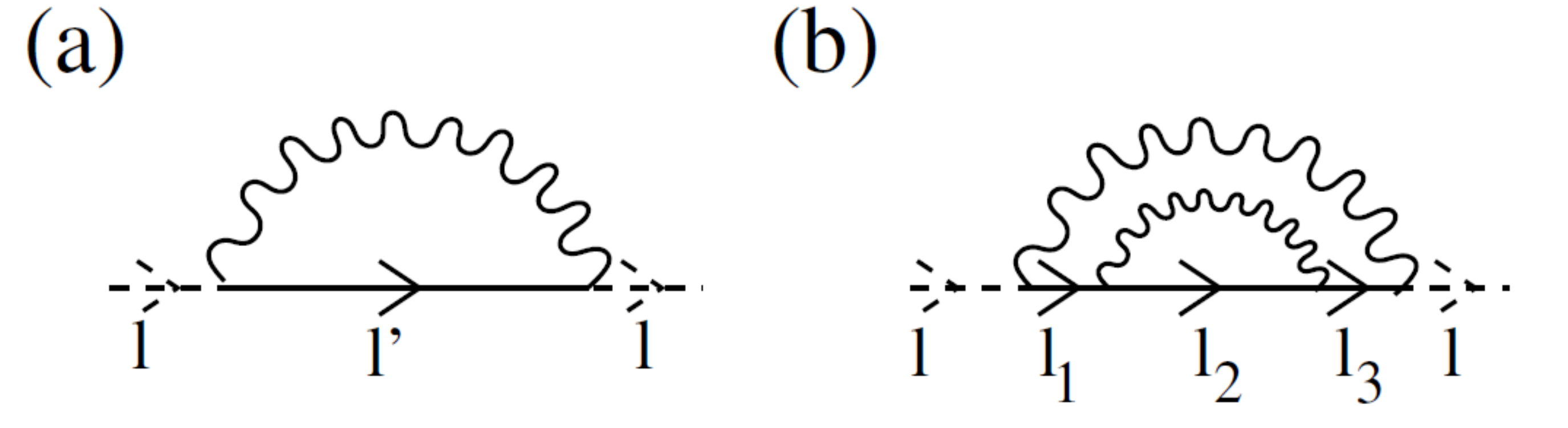}
\caption{First- and second-order self-energy diagrams considered for estimating the hopping rate of the localized state $l$.}
\label{fig.SelfEnergy}
\end{figure}
\end{center}

We consider the self-energy diagrams of Fig. \ref{fig.SelfEnergy} that describe processes through which a localized state $l$ may decay to other states $l'$ by emitting or absorbing phonons. The first-order self-energy process (i.e. single phonon) shown in Fig.\ref{fig.SelfEnergy}(a) leads to the Fermi golden rule rate of Eq.\eqref{eq.FGRRate}. As argued earlier, the above rate vanishes due to the discreteness of the one-phonon bath at the absorbed energy $\D_M$. Since the first-order rate is zero, we look into the relevant self-energy diagrams at second order [Fig.\ref{fig.SelfEnergy}(b)] and consider the decay via two-phonon absorption. This gives rise to a hopping rate
\ba
 \tau_2^{-1}&\simeq&\frac{4\pi}{3}g^4 W \sum_R\left(\frac{\D_R}{\Delta_\xi}\right)^3 e^{-(R/\xi+\D_R/T)}. \label{eq.2ndorderrate}
\ea
where $W=\Delta_\xi \xi\sim t$ is the fermion band-width.
As earlier, we can obtain a VRH-like rate using the saddle point approximation, i.e.
\ba
 \tau_2^{-1}\approx {4\pi\over 3} g^4 W \tilde{T}^{3/2} e^{-2/\tilde{T}^{1/2}}, 
\ea
involving absorption of energy $\D_M\simeq \D_\xi \tilde{T}^{1/2}$ from the two-phonon bath. The above rate should be compared with the two-phonon level spacing $\d^{(2)}_\o \simeq ((\a-1)/2) \o^{\a^2}$ at $\o=\D_M$. Hence, the two-phonon level spacing exceeds the second-order Fermi-golden rule rate as $T\to 0$ due to the exponential suppression. This implies that the two-phonon assisted rate also vanishes at low temperature, like the first-order rate. The calculation of rates at orders higher than two becomes cumbersome. However, by induction, we can write down the condition for $n$-th order process as
\ba
 g^{2n} \Delta_\xi \tilde{T}^{\,n-1/2}e^{-2/\tilde{T}^{1/2}}>\tilde{\o}_n \D_M^{\,\,\a^n}=\tilde{\o}_n\D_\xi^{\a^n}\tilde{T}^{\frac{\a^n}{2}}, \label{eq.nphonon} 
\ea
where $\tilde{\o}_n\approx ((\a-1)/2)^{\a^{n-2}}$ for $n\gg 1$. It is evident that the above condition can not be satisfied for low-order processes involving a few phonons. However, since $\D_M(T)\ll 1$, the level spacing decreases very rapidly with the number $n$ of absorbed phonons. Hence, the above condition is satisfied in a process with $n\apgt  -\ln \tilde{T}/\,(2\ln\a)$. The level spacing of the corresponding $n$-phonon bath is  $\propto e^{-\varepsilon(T)/\tilde{T}^{1/2}}$, $\varepsilon(T)$ being weakly temperature dependent. We confirm this conclusion from the weak-coupling approach by going to the strong coupling limit, $g\gg t$ and $t\ll 1$, where we directly obtain the rate for the $n$-phonon process. %To this end, we estimate small-polaron hopping rate as discussed below. Some details of the calculations are given in the Supplementary Materials.

%\section{Small-polaron hopping rate: Strong-coupling approach} \label{Sec.Polaron}
In the strong coupling approach we  eliminate the electron-phonon coupling $g$ from the Hamiltonian of Eq.\eqref{Eq.Model} by means of a canonical transformation \cite{Mahan2000}, i.e.~$\bar{\mc{H}}=e^S\mc{H}e^{-S}$, where $S=\sum_{\mu,i}(g_{\mu,i}/\o_\mu)n_i(a_\mu^\dag-a_\mu)$. The transformed Hamiltonian is
\ba
  \bar{\mc{H}}&=&t\sum_{i,\d} \bar{c}^\dag_i\bar{c}_{i+\d}+\sum_i v_i n_i+\sum_\mu\o_\mu a^\dagger_\mu a_\mu-\sum_{\mu,i}\frac{g_{\mu,i}^2}{\o_\mu}n_i,\label{eq.Polaron_H1} \nn \\ 
\ea 
where $\d$ denotes the nearest neighbour sites; $\bar{c}_i=e^S c_i e^{-S}=c_i\chi_i$ is the dressed electronic operator (polaron), where the phonon operator $\chi_i=\exp{[-\sum_{\mu,i} (g_{\mu,i}/\o_\mu)(a_\mu^\dag-a_\mu)]}$. The last term in Eq.\eqref{eq.Polaron_H1} is the polaron self-energy that could be absorbed in the definition of the localized basis $\{|l\rangle\}$. We rewrite Eq.\eqref{eq.Polaron_H1} as $\bar{\mc{H}}=\bar{\mc{H}}_0+V$ where $\bar{\mc{H}}_0=\sum_l \e_l c_l^\dag c_l+\sum_\mu \o_\mu a_\mu^\dag a_\mu$ and $V=t\sum_{i,\d}(\chi^\dag_i\chi_{i+\d}-1)c^\dag_ic_{i+\d}=\sum_{ll'}(\hat{T}_{ll'}-t_{ll'})c^\dag_l c_{l'}$. The operator $\hat{T}_{ll'}=t\sum_{i,\d}\chi^\dag_i\chi_{i+\d}\psi_l^*(R_i)\psi_{l'}(R_{i+\d})$ and $t_{ll'}=t\sum_{i,\d}\psi_l^*(R_i)\psi_{l'}(R_{i+\d})$. The small residual hopping $V$ leads to both coherent and incoherent hopping processes \cite{Mahan2000}, depending on whether the phonon number changes or not, respectively, in the process. The coherent hopping can not delocalize the electronic state in 1D. Hence we only consider the incoherent process and calculate the decay rate of polaron state $l$ in lowest order in $V$ using Fermi golden rule, i.e.
\ba
 \tau^{-1}&=&2\pi\sum_{\{n_\mu\}}\rho_{ph}\sum_{f\neq i}|\langle f|V|i\rangle|^2\delta(E_f-E_i),.~~\label{eq.FGRPolaron}
\ea
Here the initial state $|i\rangle=|l,\{n_\mu\}\rangle$ with a thermal distribution of phonons $\rho_{ph}$ and the final states are $|f\rangle =|l',\{n'_\mu\}\rangle$ such that $E_f-E_i=\e_{l'}-\e_l+\sum_\mu (n_\mu'-n_\mu)\o_\mu$; $\rho_{ph}=e^{-\b\sum_\mu n_\mu\o_\mu}/\mc{Z}_{ph}$ with $\mc{Z}_{ph}=\prod_\mu (1-e^{-\b\o_\mu})^{-1}$ ($\b=1/T$). Unlike the weak-coupling case, here the lowest order rate already contains arbitrary high order phonon processes. The rate can be obtained as
\ba
\tau^{-1}\approx 2\pi\frac{t^2}{\xi}\sum_{l'} e^{-|R_{l'}-R_l|/\xi} \mc{S}(\e_{l'}-\e_l). \label{eq.PolaronRate}
\ea
The quantity 
\ba
\mc{S}(\o)=\int_{-\infty}^\infty ds e^{-i\o s} \langle \chi_i^\dag(s)\chi_{i+\d}(s)\chi_{i+\d}^\dag(0)\chi_i(0)\rangle \label{eq.PolaronBath}
\ea
constitutes the effective bath DOS, where $\chi_i(s)=e^{i\mc{H}_{ph}s}\chi_i e^{-i\mc{H}_{ph}s}$. To obtain a non-zero rate we should compare the level spacing of $S(\o)$ with the rate $\tau^{-1}$. To this end, we expand the rate in Eq.\eqref{eq.PolaronRate} as a series in number ($n$) of phonons involved \cite{Emin1974} and compare to the level spacing of $n$-phonon process. The hopping rate associated with phonon absorptions is 
\ba
\tau^{-1}\approx\tilde{t}^{\,2}\sum_R e^{-(R/\xi+\D_\xi/2T)} \sum_{n=1}^\infty \frac{g^{2n}}{n!} I_n(\D_R),~~~\label{eq.RateExpansion} 
\ea
where $\tilde{t}^{\,2}\simeq(2\pi t^2/\xi)e^{-2S_T}$ and $e^{-2S_T}$ is a Debye-Waller factor and
\ba
I_n(\D_R)=\int_0^1 \prod_{i=1}^n d\o_i \,\o_i^{\a-1}\mr{csch}\left(\frac{\o_i}{2T}\right)\d(\D_R-\sum_i\o_i),~~\label{eq.In}
\ea
The argument of delta function contains $n$ phonon frequencies indicating the number of phonons involved in the process. Expanding the polaron hopping rate in this manner serves as an elegant way to organize the calculation of rate for high-order phonon processes. 

 We find $I_n(\D_R)\sim \D_R^{n\a-1}e^{-\D_R/2T}$ for $n$-phonon absorption (see Supplementary Materials) and the corresponding decay rate is
\ba
\tau_n^{-1}\propto \sum_R \D_R^{n\a-1}e^{-(R/\xi+\D_R/T)}
\ea 
As earlier, we obtain the saddle point rate
\ba
\tau_n^{-1}\approx C_nt\left(\frac{g^2t^\a}{\xi^\a}\right)^n\tilde{T}^{(n\a-1)/2} e^{-2/\tilde{T}^{1/2}}.  
\ea
$C_n$ is a numerical coefficient that depends on $\a$ and we have used the fact that $\D_\xi\simeq t/\xi$. Again, as in Eq.\eqref{eq.nphonon}, comparing with the $n$-phonon level spacing, non-zero rate could be obtained for $n\apgt n_c(T)\simeq -\ln\tilde{T}/(2\ln\a)$, exactly as we found in weak-coupling calculation. For $t\ll g\ll 1$, the expansion of polaron rate in Eq.\eqref{eq.RateExpansion} is controlled by the additional small parameter $g$ and the hopping rate at low temperature would be determined by the first non-vanishing one at order $\sim n_c$.  As $T\to 0$, the number of phonons, $n_c$, involved in the decay process diverges leading to a modified VRH rate law
\ba
\tau^{-1}\sim \tilde{T}^{\frac{\a}{4\ln\a}\ln(1/\tilde{T})} e^{-2/\sqrt{\tilde{T}}},
\ea
where the pre-exponential power law acquires a singular exponent as $T\to 0$. 

%\section{Discussions} \label{Sec.Discussions}
We can ask about the VRH law in higher dimensions, e.g.~in 2D, for a similar model of electron coupled to phonons in a random solid. Using Eq.\eqref{eq.FGRRate}, the first-order VRH rate is  $\tau^{-1}\sim \tilde{T}^{2/3} \exp{(-A_2/\tilde{T}^{1/3})}$. In 2D, the phonon localization length \cite{John1983,Gurarie2003} diverges as $\ell_\o\simeq \exp{(1/\o^2)}$ as $\o\to 0$, for weak disorder. This leads to a one-phonon level spacing $\d_{\D_M}\sim \tilde{T}^{-2/3}\exp{(-\varepsilon_2/\tilde{T}^{4/3})}$, at the optimal energy $\D_M\approx \D_\xi\tilde{T}^{2/3}$ with $\varepsilon_2=\D_\xi^{-2}$. As a result the inequality $\tau^{-1}>\d_{\D_M}$ is satisfied in the low temperature limit giving rise to a non-zero hopping rate in 2D, even for the lowest order phonon process. This, however, leaves open the fate of VRH law in case of strongly disordered harmonic solid in 2D. The asymptotic behavior of phonon localization length is not known in the latter case.

In conclusion, we have shown that in a one dimensional system of localized particles coupled to marginally localized phonons the usual VRH hopping mechanism is ineffective due to discreteness of the phonon bath at the relevant energy. 
Instead, the particles delocalize through a many-body process involving a large number of phonons, which diverges at low temperatures as $-\log(T/T_0)$. The many-body character of the delocalization results in a singular modification to Mott's law of variable range hopping. Hence a system of particles coupled to phonons cannot be many-body localized, however transport and thermalization are slowed down significantly.

We note that our analysis and results are not applicable to marginally localized baths, whose low frequency limit is described by an infinite randomness fixed point \cite{Dasgupta1979,Fisher1992}. Examples of such systems include the $XX$ spin chain, the critical Ising model with random interactions and transverse fields and the critical Majorana chain, i.e.~Kitaev model at the transition between the topological and trivial phases \cite{Kitaev2001}.
Like the phonons of the random harmonic chain the localization length in these systems diverges upon approaching zero energy. However the wave-functions of the low energy excitations are very different from those of the phonons. Specifically these excitations correspond to ``singlets" formed between sites at increasing distances. Hence the excitation is essentially bi-local, and  the matrix element for coupling an electron to it would be exponentially small in the distance of the electron to one of the two sites forming the singlet even if the electron is within the ``localization length" of the excitation. Another way to view this is that the electron couples to asymptotically delocalized wavefunctions albeit ones having a vanishing fractal dimension. It is an interesting open question to establish the stability, or lack thereof, of the localized state under these conditions. 

\noindent{\em Acknowledgements --} We thank Dmitry Abanin and Gil Refael  for helpful discussions. This research was supported in part by the ERC synergy grant UQUAM, and the ISF under grant 1594-11.

\begin{widetext}
\section{Supplementary Material}
 
\renewcommand{\thesection}{S\arabic{section}}    
\renewcommand{\thefigure}{S\arabic{figure}}
\renewcommand{\theequation}{S\arabic{equation}} 
%\makeatletter
%\renewcommand\@biblabel[1]{[S#1]}
%%\makeatletter
%%\renewcommand\citeform[1]{S#1}
%\makeatother

%\counterwithin{figure}{section}

\setcounter{figure}{0}
\setcounter{equation}{0}
%\setcounter{section}{0}
%\tableofcontents
In this Supplementary Material, we give some of the details of the calculations reported in the main paper. 
\subsection{1. Phonon density of states and level spacing} \label{App.PhononDOS}
We first discuss the calculation of level spacing for single- and multi-phonon processes. The phonon modes that have finite overlap at $R\simeq R_l$ contribute to the single-phonon hopping rate [Eq.(1)] of localize state centred at $R_l$. The number (per unit energy) of such phonon modes can be estimated as:
\ba
N_1(\o)&\approx &\sum_\mu e^{-|R_l-R_\mu|/\ell_\mu}\d(\o-\o_\mu)\approx \ell_\o^d\nu(\o) \nn
\ea
The relevant level spacing is $\d_\o^{(1)}\approx 1/N_1(\o)\approx 1/\nu(\o)\ell_\o^d$. In 1D, $\d_\o^{(1)}\approx \o^\a$. Similarly, for the two-phonon absorption process
\ba
N_2(\o)&\approx& \sum_{\mu_1,\mu_2}e^{-|R_l-R_\mu|/\ell_\mu}e^{-|R_l-R_{\mu'}|/\ell_{\mu'}}\d(\o-\o_\mu-\o_{\mu'})\simeq \int d\o_1 d\o_2 \ell_{\o_1}\ell_{\o_2}\nu(\o_1)\nu(\o_2)\d(\o-\o_1-\o_2) \nn 
\ea
The lower limit of the frequencies in the above integral needs to be treated with care, otherwise the integral is apparently divergent. The lower limit is essentially determined by the discreteness of the one-phonon DOS. The maximum possible value of one of the phonon frequencies, say $\o_1$, is $\sim \o-\d_\o^{(1)}$. Hence, the delta function in the integrand constrains the minimum possible value of $\o_2$ to $\d_\o^{(1)}$. As a result
\ba
N_2(\o)\approx \int_{\d_\o^{(1)}}^{\o-\d_\o^{(1)}}d\o'\frac{1}{(\o-\o')^\a}\frac{1}{\o'^\a} \label{eq.2PhononDOS}
\ea
For $\a>1$, the integral is dominated by the lower limit for $\d_\o^{(1)}\ll\o$. This gives $N_2(\o)\propto \o^{-\a^2}$ and the two-phonon level spacing $\d_\o^{(2)}\propto \o^{\a^2}$. This could be confirmed by evaluating the integral in 
Eq.\eqref{eq.2PhononDOS} more accurately through a series expansion around the lower limit, i.e.
\ba
 \int_{\d_\o^{(1)}}^{\o-\d_\o^{(1)}}d\o'\frac{1}{(\o-\o')^\a}\frac{1}{\o'^\a}=\frac{2}{\o^\a}\sum_{n=0}^\infty \frac{(-1)^n}{\o^n}\frac{(-\a)!}{n!(-\a-n)!}\int_{\d_\o^{(1)}}^{\frac{\o}{2}} d\o' \o'^{n-\a}\simeq\frac{2}{\a-1}\o^{-\a^2},
\ea
where we keep only the most singular term as $\o\to 0$. The two-phonon level spacing, $\d_\o^{(2)}\approx 1/N_2(\o)$ is obtained as
\ba
\d_\o^{(2)}\approx \tilde{\o}_2\o^{\a^2}.\nn  
\ea
The frequency scale $\tilde{\o}_2=((\a-1)/2)$.
Similarly, the three-phonon DOS can be written as
\ba
N_3(\o)&=& \int_{\o_1>\d_\o^{(2)},\o_2>\d_\o^{(1)}} d\o_1 d\o_2 N_1(\o_1)N_2(\o_2)\d(\o-(\o_1+\o_2))\nn
\ea
leading to $\d_\o^{(3)}\approx \tilde{\o}_3\o^{\a^3}$ with $\tilde{\o}_3=\frac{\a-1}{2}(\frac{1}{\a^2-1}+\frac{1}{2}(\frac{2}{\a-1})^\a)^{-1}$. The $n$-phonon level spacing can be obtained recursively and  we get $\d_\o^{(n)}\approx \tilde{\o}_n \o^{\a^n}$, where $\tilde{\o}_n=((\a-1)/2)^{\a^{n-2}}$ for $n\gg 1$.

 The above results for the level spacings are only applicable to $\a>1$. The limit $\a=1$ is special and only asymptotically reached, e.g.~for the case of phonons of a superfluid, at the critical point corresponding to the superfluid-insulator transition \cite{Gurarie2008_S}. The multi-phonon level spacings in this case can be evaluated exactly \footnote{Neglecting a logarithmic correction to scaling \cite{Gurarie2008_S}, i.e. $\ell_\o\sim \ln^2(\o)/\o$.}. For the two-phonon process, from Eq.\eqref{eq.2PhononDOS}, it is straightforward to obtain $N_2^\a(\o)\propto (1/\o)$ and  $\d_\o^{(2)}\propto \o$. Similarly, for the $n$-th order $\d_\o^{(n)}\propto \o$, implying that the rapid decrease of level spacing that leads to the eventual thermalization via arbitrary high order phonon process in case of $\a>1$, is absent at $\a=1$.

\subsection{2. Electron-phonon coupling} \label{App.el-ph_coupling}
The el-ph coupling term in Eq.(4) (main text) could be obtained by expanding (short-range) electron-ion potential $V_{ei}$ in terms of displacement $x_i$'s 
\ba
\mc{H}_{el-ph}\sim \sum_i n_i V_{ei}(R_{i+1}-R_i)(x_{i+1}-x_i) 
\ea
This leads to 
\ba
g_{\mu,i}\sim \frac{1}{\sqrt{\o_\mu}}\sum_i V_{ei}(R_{i+1}-R_i)\partial_i\phi_\mu(R_i)
\ea
where 
\ba
\partial_i\phi_\mu(R_i)=\frac{\phi_\mu(R_{i+1})}{\sqrt{m_{i+1}}}-\frac{\phi_\mu(R_i)}{\sqrt{m_i}}\sim \frac{e^{-|R_i-R_\mu|/2\ell_\mu}}{\l_\mu \ell_\mu^{1/2}},
\ea
 with $\lambda_\mu\simeq 1/\o_\mu$ the phonon wavelength, giving rise to the form of el-ph coupling in Eq.(5) of the main paper.

\subsection{3. Perturbative hopping rate: Weak-coupling calculation} \label{App.2ndorderRate}
\begin{center}
\begin{figure}
\includegraphics[width=0.4\textwidth]{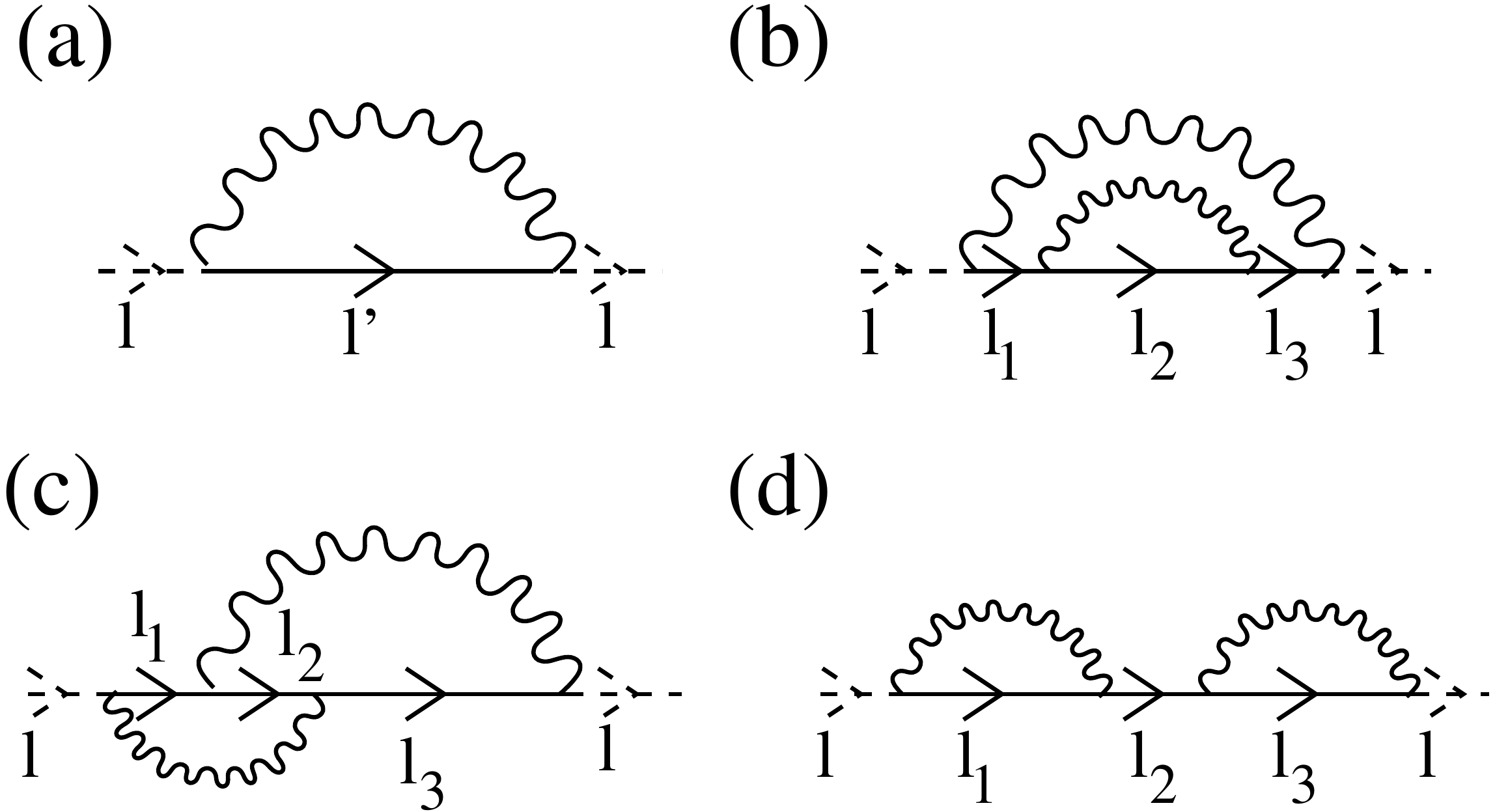}
\caption{First- and second-order self-energy diagrams. The solid line with the arrow corresponds to electron propagator and the curly line to the phonon propagator.}
\label{fig.SelfEnergy_S}
\end{figure}
\end{center}

The single-phonon process in Fig.\ref{fig.SelfEnergy_S}(a) gives rise to the first-order self-energy 
\ba
\S_{ll}^{(1)}(i\o_n)=-T\sum_{\mu,l',\O}|g_{ll',\mu}|^2D_\mu(i\O)G^0_{l'}(i\o-i\O), 
\ea
where $g_{ll',\mu}=\sum_i g_{\mu,i}\psi_l^*(R_i)\psi_{l'}(R_i)=g_{l'l,\mu}^*$ and $D_\mu(i\O_m)=-2\o_\mu/(\O_m^2+\o_\mu^2)$ is the phonon- and $G^0_l(i\o_n)=1/(i\o_n-\e_l)$ is the bare electron-propagators; $\O_m$ and $\o_n$ are bosonic and fermionic Matsubara frequencies, respectively. The rate $\tau_1^{-1}=-2\mr{Im}\S_{ll}^{(1)}(\e_l+i0^+)$ at $\mc{O}(g^2)$ via single-phonon absorption ($\e_{l'}>\e_l$) is
\ba
\tau_1^{-1}&=&2\pi\sum_{l',\mu}|g_{\a\b,\mu}|^2\d(\e_l-\e_{l'}-\o_\mu)n_\mr{B}(\o_\mu). 
\ea
[$n_\mr{B}(\o)=1/(e^{\o/T}-1)$]. By converting the sum, $\sum_\mu\approx \int_0^1d\o \nu(\o)\int dR_\mu$, and using $\int dR_\mu e^{-|R_l-R_\mu|/\ell_\o}\approx \ell_\mu$ we obtain 
\ba
\tau_1^{-1}&=&2\pi g^2\sum_R \D_R e^{-R/\xi} n_\mr{B}(\D_R), \label{eq.1storderRate}
\ea
where $R=|R_l-R_{l'}|$. In the limit $T\ll\D_R$, the above leads to the golden rule rate of Eq.(1).

The second-order self-energy diagram of Fig.\ref{fig.SelfEnergy_S}(b) gives
%\begin{widetext}  
\ba 
\Sigma_{ll}^{(2)}(i\o)&=&T^2\sum_{l_1l_2l_3,\Omega_1,\omega_2,\mu_1,\mu_2}g_{ll_1,\mu_1}g_{l_1l_2,\mu_2}g_{l_2l_3,\mu_2}g_{l_3l_,\mu_1} D_{\mu_1}(i\Omega_1)D_{\mu_2}(i\Omega_2)G_{l_1}^0(i\o-i\Omega_1)G_{l_2}^0(i\o -i\Omega_1-i\Omega_2) G_{l_3}^0(i\o-i\Omega_2)\nn 
\ea
As $g_{ll',\mu}=\sum_i g_{\mu,i}\psi_l^*(R_i)\psi_{l_1}(R_i)\sim e^{-|R_l-R_\mu|/2\xi}e^{-|R_l-R_{l'}|/2\xi}e^{-i\th_{ll',\mu}}$, the product of electron phonon couplings in the above gives rise to random phases for $l_3\neq l_1$, i.e.
\ba
g_{ll_1,\mu_1}g_{l_1l_2,\mu_2}g_{l_2l_3,\mu_2}g_{l_3l,\mu_1}=|g_{ll_1,\mu_1}||g_{l_1l_2,\mu_2}||g_{l_2l_3,\mu_2}||g_{l_3l,\mu_1}| e^{-i(\theta_{ll_1,\mu_1}+\theta_{l_1l_2,\mu_2}+\theta_{l_2l_3,\mu_2}+\theta_{l_3l,\mu_1})}\nn
\ea
Hence the main contribution comes for $l_3=l_1$. The self-energy due to two-phonon absorption is
\ba
\Sigma_{ll}^{(2)}(i\o)=T^2\sum_{l_1l_2,\Omega_1,\Omega_2,\mu_1,\mu_2} \frac{|g_{ll_1,\mu_1}|^2|g_{l_1l_2,\mu_2}|^2}{(i\Omega_1+\omega_{\mu_1})(i\Omega_2+\omega_{\mu_2})(i\o-i\Omega_1-\epsilon_{l_1})^2(i\o-i\Omega_1-i\Omega_2-\epsilon_{l_2})},
\ea
and the corresponding rate
\ba
&&\tau_2^{-1}\sim \sum_{l_1l_2,\mu\nu} \mr{Re}\frac{|g_{ll_1,\mu}|^2|g_{l_1l_2,\nu}|^2}{(\e_l-\e_{l_1}+\o_\mu+i0^+)^2}\delta(\e_l-\e_{l_2}+\o_\mu+\o_\nu)
\label{eq.2ndorderrate_S}
\ea
Note that the second order rate also has additional contribution from processes which basically involve one-phonon DOS, e.g.~terms $\sim \delta(\e_l-\e_{l'}+\o_{\mu})$, essentially like first-order rate. However, since the single-phonon absorption does not have `sufficiently' continuous DOS to give rise to non-zero rate even at $\mc{O}(g^2)$, we do not consider these processes at $\mc{O}(g^4)$. 

The rate in Eq.\eqref{eq.2ndorderrate_S} can be written as
\ba
\tau_2^{-1}&\simeq&2\pi g^4\int_{\o_1,\o_2}\sum_{l_1l_2} \mr{Re}\frac{\o_1\o_2 n_\mr{B}(\o_1)n_\mr{B}(\o_2)}{(\e_l-\e_{l_1}+\o_1+i0^+)^2} e^{-|R_l-R_{l_1}|/\xi}e^{-|R_{l_2}-R_{l_1}|/\xi} \delta(\e_l-\e_{l_2}+\o_{\mu_1}+\o_{\mu_2})
\ea
%\end{widetext}
We carry out the sum over $l_1$ for $\o_1\ll \D_\xi$,
\ba
&&\tau_2^{-1}\simeq 8\pi\frac{g^4\xi}{\D_\xi^2}\sum_R e^{-\frac{R}{\xi}}\int_0^{\D_R} d\o \o(\D_R-\o)n_\mr{B}(\o)n_\mr{B}(\D_R-\o) \nn
\ea
Here $R=|R_l-R_{l_2}|$ and $\D_R=|\e_{l_2}-\e_l|$. For $T\ll \D_R$ the frequency ranges $0<\o\aplt T$ and $0<\D_R-\o\aplt T$ does not contribute substantially to the integral. Hence we approximate $n_\mr{B}(\o)n_\mr{R}(\D_R-\o)\approx e^{-\D_R/T}$ to obtain Eq.(6). 

 There are two other types of diagram at second order, as shown in Figs.\ref{fig.SelfEnergy_S}(c) and (d). The `crossing' diagrams of Fig.\ref{fig.SelfEnergy_S}(c) corresponds to vertex correction plus real decay process involving a single-phonon. This is effectively equivalent to the first order diagram of Fig.\ref{fig.SelfEnergy_S}(a) with a renormalized el-ph coupling. The discreteness of the single-phonon spectra makes the corresponding decay rate zero, as in the case of first-order process. Similar argument can be given for all higher order crossing diagrams, i.e. if the decay rate is zero at a given order, then the crossing diagrams at the next higher order also can not contribute to the decay. Hence we only consider non-crossing diagrams to find out the first non-vanishing rate.
 
Due to the lack of translational invariance, the self-energy has off-diagonal components $\S_{ll'}$ for $l\neq l'$. As a result
\ba
G_{ll'}(\o)=G^0_l(\o)\delta_{ll'}+G^0_l(\o)\S_{ll_1}(\o)G_{l_1l'}(\o) 
\ea
(repeated indices summed over). With some algebraic manipulation one can write 
\ba
G_{ll}(\o)=\frac{1}{(G^0_l(\o))^{-1}-S_l(\o)} 
\ea
where $S_l=\S_{ll}+\S_{ll'}G^0_{l'}\S_{l'l}+\dots$ . The decay rate is given by $-\mr{Im}S_l(\o)$ \cite{Fleishman1980}. Hence the off-diagonal self-energy at $\mc{O}(g^2)$ contributes to $\mc{O}(g^4)$ decay rate, e.g. as in Fig.\ref{fig.SelfEnergy_S}(d) for the second order rate. However, the contribution of off-diagonal self-energy to the rate essentially involves single-phonon processes. As earlier, if the first-order rate is zero then these off-diagonal contributions in the second order are also zero.

\subsection{4. Small polaron hopping rate: Strong-coupling calculation} \label{App.SmallPolaron}
%\begin{widetext}
The polaron rate of Eq.(11) can be obtained from Eq.(10) by using $E_f-E_i=\e_{l'}-\e_l+\sum_\mu (n_\mu'-n_\mu)\o_\mu$ and
\ba
&&2\pi\sum_{f\neq i}|\langle f|V|i\rangle|^2\delta(E_f-E_i)=\sum_{f\neq i}\int_{-\infty}^\infty ds e^{-i(E_f-E_i)s}|\langle i|c^\dag_l(\hat{T}^\dag_{l'l}-t^*_{l'l})c_{l'}|f\rangle|^2\nn\\
&&=\sum_{l'\neq l}\int ds e^{-i(\e_{l'}-\e_l)s}\left[\langle n_\mu|\hat{T}^\dag_{l'l}(s)\hat{T}_{l'l}|n_\mu\rangle-|\langle n_\mu|\hat{T}_{l'l}|n_\mu\rangle|^2\right]\label{eq.FGR2}
\ea
Here $\hat{T}^\dag_{ll'}(s)=e^{i\mc{H}_{ph}s}\hat{T}^\dag_{ll'}e^{-i\mc{H}_{ph}s}$. The last term in Eq.\eqref{eq.FGR2} does not contribute as $\e_{l'}\neq \e_l$. The transition rate is
\ba
\tau^{-1}&=& t^2\sum_{l'\neq l}\int ds e^{-i(\e_{l'}-\e_l)s}\sum_{i,j,\d,\d'}\psi_{l'}(R_i)\psi^*_{l'}(R_j)\psi^*_l(R_{i+\d})
\psi_l(R_{j+\d'})\langle\chi^\dag_i(s)\chi_{i+\d}(s)\chi^\dag_j(0)
\chi_{j+\d'}(0)\rangle \label{eq.FGR3}
\ea
$\langle ..\rangle$ denotes thermal average over phonons. The wavefunction overlap  
\ba
\psi_{l'}(R_i)\psi^*_{l'}(R_j)\psi^*_l(R_{i+\d})
\psi_l(R_{j+\d'})\sim e^{i(\th_{l'}(R_i)-\th_{l'}(R_j))} e^{-i(\th_l(R_{i+\d})-\th_l(R_{j+\d'}))}
\ea
% \end{widetext} 
gives rise to random phases for $R_i\neq R_j$ and $R_{i+\d}\neq R_{j+\d'}$. Hence the main contribution comes from $R_i=R_j$, $\d=\d'$ leading to Eq.(11). The correlation function $\langle\chi^\dag_i(s)\chi_{i+\d}(s)\chi^\dag_i(0)
\chi_{i+\d}(0)\rangle$ can be evaluated following standard procedure \cite{Mahan2000_S} as
%\begin{widetext}
\ba
\langle\chi^\dag_i(s)\chi_{i+\d}(s)\chi^\dag_i(0)
\chi_{i+\d}(0)\rangle=e^{-2S_T}\exp{\left({\sum_\mu\frac{(g_{\mu,i}-g_{\mu,i+\d})^2}{\o_\mu^2}\mr{csch}\left(\frac{\o_\mu}{2T}\right)\cos\left(\o_\mu(s+i/2T)\right)}\right)} \label{eq.FGR5}
\ea
%\end{widetext}
The exponent in the Debye-Waller factor $e^{-2S_T}$ is
\ba
2S_T=&&\sum_\mu\frac{1}{\o_\mu^2}(g_{\mu,i}-g_{\mu,i+\d})^2
\coth\left(\frac{\o_\mu}{2T}\right)\approx 2g^2\int_0^1 d\o \o^{\a-1}\coth\left(\frac{\o}{2T}\right)\nn
\ea
and 
\ba
&&\sum_\mu\frac{(g_{\mu,i}-g_{\mu,i+\d})^2}{\o_\mu^2}
\mr{csch}\left(\frac{\o_\mu}{2T}\right)\cos\left(\o_\mu(s+\frac{i}{2T})\right)\approx 2g^2\int_0^1 d\o \o^{\a-1}\mr{csch}\left(\frac{\o}{2T}\right)\cos\left(\o(s+\frac{i}{2T})\right)\nn
\ea
Using the above we evaluate $\mc{S}(\o)$ in Eq.(11). Finally, the polaron hopping rate can be obtained as:
\ba 
\tau^{-1}&\approx& \frac{t^2}{\xi}\sum_R e^{-R/\xi}~~e^{-2S_T}e^{-\D_R/2T}\int_{-\infty}^\infty ds e^{-is\D_R }e^{2g^2\int_0^1 d\o \o^{\a-1}\mr{csch}(\o/2T)\cos(\o (s+i/2T))}\nn
\ea
where $R=|R_{l'}-R_l|$ and $\D_R=\e_{l'}-\e_l$. This could be expanded in the form of the series in Eq.(13).

For one-phonon process it is straightforward to obtain,  $I_1(|\e|)=|\e|^{\a-1}\mr{csch}\left(\frac{|\e|}{2T}\right)\simeq 2|\e|^{\a-1}e^{-|\e|/2T}$ for $|\e|\gg T$, from Eq.(14). The two-phonon absorption ($\e>0$) leads to
\ba
I_2(\e)=  \int _0^\e d\o \left[\o(\e-\o)\right]^{\a-1}\mr{csch}\left(\frac{\o}{2T}\right)\mr{csch}\left(\frac{\e-\o}{2T}\right) \nn
\ea
Since $\a>1$ and $\mr{csch}(x)\approx 1/x$ as $x\to 0$, the frequency ranges $0<\o\aplt T$ and $0<(\e-\o)\aplt T$, i.e. the neighbourhood of lower and upper limits, do not contribute substantially to the above integral for $\e\gg T$, the main contribution comes from $\o\sim \e/2$. As a result, we can approximate $\mr{csch}(\o/2T)\approx 2 e^{-\o/2T}$ and $\mr{csch}((\e-\o)/2T)\approx 2 e^{-(\e-\o)/2T}$ so that
$I_2(\e)\approx 4 e^{-\e/2T}\int_0^\e d\o \o^{\a-1}(\e-\o)^{\a-1}\sim \e^{2\a-1} e^{-\e/2T}$. From Eq.(15), the quantity $I_3(\e)$ for the three-phonon absorption can be rewritten and evaluated as:
\ba
I_3(\e)=\int_0^\e d\o \o^{\a-1}\mr{csch}\left(\frac{\o}{2T}\right)I_2(\e-\o)\sim \e^{3\a-1}e^{-\e/2T}.
\ea
By induction, the $n$-phonon absorption leads to $I_n(\e)\sim \e^{n\a-1}e^{-\e/2T}$.

\end{widetext}

\end{document}